\documentclass[final,3p,times,twocolumn]{elsarticle}
 \biboptions{comma,sort&compress}
\usepackage{here}
 \usepackage{graphicx}
  \usepackage{epsfig}

\usepackage{amsthm}
\usepackage{amsmath}
\usepackage{amssymb}

\journal{Nuc. Phys. (Proc. Suppl.)}

\begin{document}

\begin{frontmatter}



\title{ChPT calculations of pion formfactors}

 \author[label1]{Karol Kampf}
  \address[label1]{Institute of Particle and Nuclear Physics,
Faculty of Mathematics and Physics,\\
Charles University, V Hole\v{s}ovi\v{c}k\'ach 2, 18000 Prague, Czech Republic}
\ead{karol.kampf@mff.cuni.cz}


\begin{abstract}
\noindent
An overview on chiral perturbation theory calculations of form factors is
presented.
The main focus is given on the form factors related to the lightest meson,
pion, namely:
pion decay constant, pion vector and scalar form factor, radiative pion
decay and transition
form factor.  A pure calculation within the effective theory can be extended
using further methods, as
resonance chiral theory and leading logarithm calculations.

\end{abstract}


\end{frontmatter}


\section{Introduction}
\noindent
The formfactors of quantum chromodynamics (QCD) are well defined objects which
can be studied both theoretically and experimentally. We will focus on several
basic quantities which are connected with $\pi$ meson and summarize
basic status of their theoretical calculations mainly at low energies, i.e. at
the domain of chiral perturbation theory (ChPT). The formfactors connected for example with kaons will not be considered here, but one should note that they play also important role in connection with ChPT (e.g. $K_{\ell4}$).


\section{Pion decay constant}
\noindent
The most simplest formfactor, pion decay constant, is defined in QCD via the
coupling of axial current and pion as
\begin{equation}
 \langle 0| A_\mu^a(x) | \pi^b(p)\rangle = i \delta^{ab} F_\pi p_\mu {\rm
e}^{-ipx}\,.
\end{equation}
As the pion is real, $p^2 = m_\pi^2$, the momentum dependence is trivial and
$F_\pi$ is a constant. This is a reason why it is usually not referred as
formfactor in the literature (on recent review see e.g. \cite{Gasser:2010wz} and
references therein).
It is a fundamental order parameter of the spontaneous symmetry breaking of
$SU(N_f)_L\times SU(N_f)_R$ to $SU(N_f)_V$ of QCD ($N_f$ represents number of
light quark flavours, 2 or 3 for real QCD). Its value can be set from the
$\pi_{\ell2}$ decay using Marciano and Sirlin formula for radiative corrections
\cite{Marciano:1993sh}. Updated by virtual photons \cite{virphot} and $V_{ud}$
value \cite{Beringer:1900zz} one can obtain \cite{Kampf:2009tk}
\begin{equation}\label{Fpi1}
 F_\pi = 92.215 \pm 0.0625\, \text{MeV}\,.
\end{equation}
In pure QCD, $F_{\pi^0}$ and $F_{\pi^\pm}$ difference is NNLO effect and this
was evaluated in \cite{Kampf:2009tk} and found to be indeed very small. We use
this fact in order to set the pion decay constant form $\pi^0$ lifetime. Using
the NNLO calculation within ChPT of $\pi^0\to \gamma\gamma$ decay
\cite{Kampf:2009tk} subtracting QED corrections one can arrive to
\begin{equation}\label{Fpi2}
 F_{\pi^0} = 93.85 \pm 1.3 \text{ (exp.) } \pm 0.6 \text{ (theory) MeV}\,.
\end{equation}
As an experimental input the PrimEx measurement was used
\cite{Larin:2010kq}. We can see that the precision obtained here cannot still
compete with the precision obtained using charged pion decay. However new
experimental activity (e.g. PrimEx2, KLOE-II) can improve the experimental
error. On the theory side there are also possible improvements foreseen. One of
them, the full calculation of the $\eta\to\gamma\gamma$ decay will
be valuable \cite{Bijnens:2010pa}, as well as a better estimation of the
value of the isospin
breaking coefficient $\sim (m_d-m_u)$. What is important to stress at this
moment is that possible tension between these two values~(\ref{Fpi1})
and~(\ref{Fpi2}) can be attributed to new physics: $F_\pi$, determined from
the weak decay of the $\pi^+$ assumed the validity of the standard model. A possible deviation from it via right-handed currents was
opened in \cite{Bernard:2007cf}.


\section{Electromagnetic formfactor of charged pion}
\noindent
Vector or more precisely electromagnetic formfactor of the charged pion,
$F^{\pi}_V$ is defined by
\begin{equation}
 \langle \pi^+(p_f)| j_\mu^\text{elm} | \pi^+(p_i)\rangle = (p_f+p_i)_\mu
F_V^\pi [(p_f-p_i)^2]\,.
\end{equation}
Its calculation within ChPT up to two-loop level can be found
in~\cite{BijnensFV} and using dispersive treatment in \cite{Gasser:1990bv}.
Data within the validity of ChPT were taken so far mainly from a space like
region (cf.\cite{BijnensFV}). New measurements in a time-like region almost down
to the di-pion threshold (at KLOE10 \cite{Ambrosino:2010bv}) urge us to
answer the question of validity of ChPT more precisely. For this we will turn
to the calculations of the leading logarithms.

\begin{figure}[hbt]
\centerline{\includegraphics[width=10.5cm]{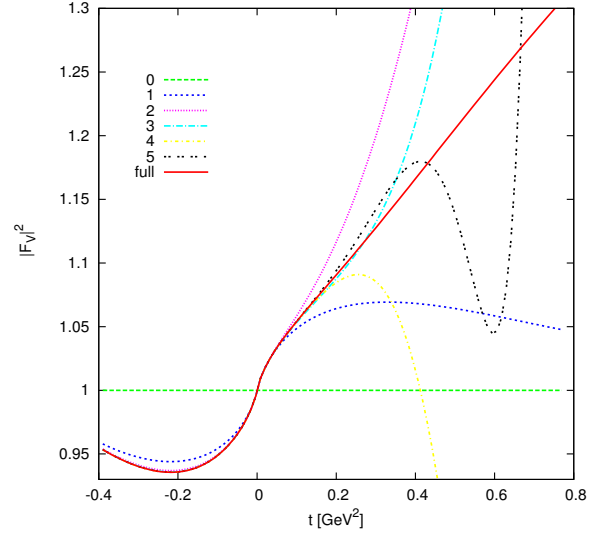}}
\caption{\scriptsize The leading-logarithms for normalized $F_V$ in massless
and large $N$ limit in $O(N=3)$ model.}
\label{fig1}
\end{figure}

{\it Leading logarithms} (LL) are logarithms with highest possible power at the
given order. Similarly as in the renormalizable theory they can be calculated
using only one-loop diagrams \cite{LLproof}. In the renormalizable theory
their summation has an important phenomenological consequence: the running coupling
constant. In effective theory, as ChPT, the LL coefficients are given only by
the form of the leading-order Lagrangian. They are thus parameter-free and without
further knowledge of low-energy constants can be used as a rough estimate of the
given order. However, the general method for their summation is not known and
thus at the moment we must rely only on some simplify cases where it was
possible. LL were calculated up to the fifth order in the massive $O(N)$ model
(for $N=3$ it is equivalent to two-flavour ChPT) in
\cite{Bijnens:2010xgBijnens:2012hf}. In the massless and large $N$ limit it is
indeed possible to resum all LL, the closed form is (cf. also
\cite{Kivel:2009az}, but mind the sign)
\begin{equation}
F_V^{0NLN}(t) =
1+\frac1N +\frac{4}{K_tN^2} \bigg[
1-\Big(1+\frac{2}{K_tN}\Big) \log \Big(1 + \frac{K_tN}{2} \Big)
\bigg]\,,
\end{equation}
with
\begin{equation}
 K_t \equiv \frac{t}{16\pi^2 F^2} \log\Big(-\frac{\mu^2}{t} \Big) \, .
\end{equation}
The calculated LL together with the resummed function is depicted in
Fig.~\ref{fig1}.
It is clear that convergence in the time-like region is already  problematic
not far above the threshold. It also shows how important is a resum function (at
least in the studied limits). Let us note that even LL are very important in
studying the convergence, the actual numerical value is still dominated by the
large higher-order coefficients \cite{BijnensFV}.


\section{Scalar formfactor}
\noindent
The definition reads
\begin{equation}
  F_S^\pi (t\equiv(p-q)^2) = \langle \pi^0(q) | \bar uu + \bar dd| \pi^0(p)
\rangle
\end{equation}
Its calculation within ChPT exists up to next-to-next-to-leading order
in~\cite{Bijnens:2003xg}. The fact that
this quantity cannot be practically measured today can appear as a problem. However, using $\pi\pi$
phase shifts and the dispersive treatment \cite{Moussallam:1999aq} we can study
its energy dependence. Writing
\begin{equation}
F_S^\pi(t) = F_S^\pi(0) \Bigl( 1+\frac16\langle r^2 \rangle_S^\pi t + c_S^\pi
t^2 +\ldots \Bigr)\,,
\end{equation}
one would obtain
\begin{equation}
\langle r^2 \rangle_S^\pi = 0.61\pm0.04\;\text{fm}^2,\qquad c_S^\pi =
11\pm2\;\text{GeV}^{-4}\,.
\end{equation}
These values were recently used in the new global fit of low energy constants
of the 3-flavour ChPT \cite{Bijnens:2011tb}.

\section{Radiative pion decay}
\noindent
The pion decay $\pi^+ \to e^+ \nu \gamma$ (see e.g. works in \cite{radpi}) is
interesting in the context of the
QCD formfactors because its structure dependent part dominates over the inner
Bremsstrahlung due to the helicity suppression. The structure dependent part,
connected to the pionic structure, can be further decomposed to the vectorial
($\sim F_V$) and axial ($\sim F_A$) part. Beyond standard model one can consider
also tensor radiation part ($\sim F_T$). As there is no significant hint from
the recent measurements -- the most precise limits
are $F_T = (-0.6\pm 2.8)\times 10^{-4}$ set by the PIBETA group
\cite{Bychkov:2008ws} -- we will not consider it here. The vector part of the
$V-A$ structure, defined as ($e=1$ for simplicity)
$$
\int d^4 x {\rm e}^{i q.x} \langle 0 | T(j_\mu^\text{elm}(x) j_\nu^{V; 1-i2}(0)
| \pi^+(p)\rangle = \epsilon_{\mu\nu\alpha\beta} q^\alpha p^\beta
\frac{F_V}{m_{\pi^+}}
$$
and $\pi^0\gamma\gamma$ amplitude, defined as
\begin{equation}
 \int d^4 x {\rm e}^{i q.x} \langle 0 | T(j_\mu^\text{elm}(x)
j_\nu^\text{elm}(0)
| \pi^+(p)\rangle = \epsilon_{\mu\nu\alpha\beta} q^\alpha p^\beta
A_{\pi\gamma\gamma}
\end{equation}
can be connected employing an isospin symmetry
\begin{equation}\label{FVApigg}
\sqrt 2 \frac{F_V}{m_{\pi^+}} = A_{\pi^0\gamma\gamma}\,.
\end{equation}
The recent measurement \cite{Bychkov:2008ws} $F_V= 0.0258(17)$ is in agreement
with the value obtained either from $\pi^0\to\gamma\gamma$ decay width or
$O(p^4)$ theoretical calculation. This value was also used as an independent
determination of the neutral pion lifetime (see also \cite{Beringer:1900zz})
\begin{equation}\label{taupsi}
 \tau_{\pi^0}^\text{PSI}  = (8.5\pm1.1)\times 10^{-17} \text{ s}\,.
\end{equation}
However, one should be careful with systematic uncertainties. As we have
mentioned the connection between $F_V$ and $\pi^0\gamma\gamma$ is based on the isospin
symmetry. It also means that the value of the mass of pion
in~(\ref{FVApigg}) is just matter of convention. The dependence on the actual
value is source of roughly 50\% of the error in (\ref{taupsi}). Independently
of $\pi_{l\gamma}$ decay, the isospin-breaking corrections were found to be
very important also in the theoretical estimate of $\pi^0\to\gamma\gamma$ decay
width \cite{Kampf:2009tk}.

\section{Transition formfactor}
The transition pion-gamma-gamma (all off-shell) formfactor is a quantity
accessible via a definition of the QCD Green function of the vector-vector and
pseudoscalar currents
\begin{equation}
 \Pi_{\mu\nu}^{abc}(p,q) = \int dx dy {\rm e}^{ip.x + iq.y} \langle 0 |
T[V_\mu^a(x) V_\nu^b(y) P^c(0) ] |0\rangle\,.
\end{equation}
Using Ward identities and Lorentz and parity invariance we can extract
\begin{equation*}
  \Pi_{\mu\nu}^{abc}(p,q) = d^{a,b,c} \epsilon_{\mu\nu\alpha\beta}p^\alpha
q^\beta  \Pi(p^2,q^2;r^2\equiv(p+q)^2)\,.
\end{equation*}
The formfactors can be obtained using the LSZ. E.g. for the $\pi^0$ transition
formfactor we have (for simplicity in the chiral limit)
\begin{equation}\label{FFpigg}
 {\cal F}_{\pi^0\gamma\gamma}(p^2,q^2;r^2) = \frac23 \frac{1}{BF} r^2
\Pi(p^2,q^2;r^2)\,.
\end{equation}
The application of this object is very wide. The most important place where its
theoretical behaviour is most desired is probably the hadronic light-by-light
contribution in muonic anomalous magnetic moment. Putting a pion on shell
we can consider two regions depending on a photon virtuality: space-like
(represented e.g. by the $e^+e^-$ fusion to $\pi^0$) and time-like region
(e.g. $\pi^0\to e^+ e^- \gamma$). On the more detailed overview and literature
see the recent MesonNet workshop \cite{Czerwinski:2012ry}.

\begin{figure}[hbt]
\centerline{\includegraphics[width=8cm]{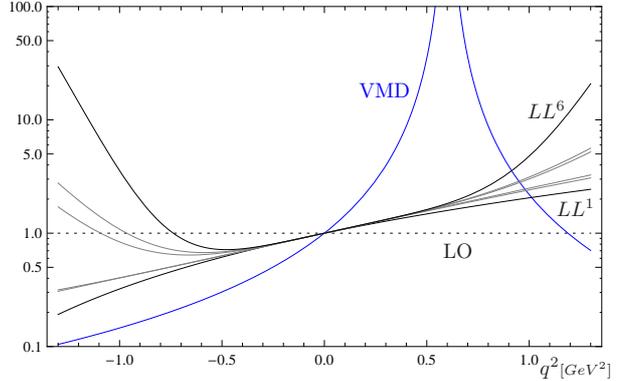}}
\caption{\scriptsize Normalized $|{\cal F}_{\pi\gamma\gamma^*}|^2$ as a function of
a virtual photon, calculated using LL only (up to 6-loop order) [e.g. $LO=1$, $LL^1 = 1+\frac{q^2}{96\pi^2F_\pi^2}\ln{\frac{\mu^2}{m_\pi^2}}$]. For comparison: vector meson dominance curve (VMD).}
\label{fig2}
\end{figure}

The area of the applicability of ChPT is demonstrated on Fig.~\ref{fig2}.
We clearly see that area is roughly $|q^2|\lesssim 0.5 \text{ GeV}$ for both
time-like and space-like $q$. Unfortunately this region is not yet very well
covered by experimental data. In the following we will discuss both regions in
more detail.


\subsection{Space-like region}
\noindent
Apart from the model-independent LL we must employ some specific model and
parameters in order to describe behaviour of the objects as defined for example
in~(\ref{FFpigg}) at low energies. In pure ChPT we have to deal with
low-energy constants. For ${\cal F}_{\pi^0\gamma\gamma}$ these are mainly
$C_7^W$ and $C_{22}^W$ of the odd-intrinsic-parity sector \cite{Bijnens:2001bb}. They
must incorporate the existence of resonances and their effect even below their
thresholds (as it is clear in Fig.~\ref{fig2}). We will be, however, still
limited with the applicability of such models strictly below these resonances.
On the other hand, one can enlarge ChPT by resonances and keep them as active
degrees of freedom. For the mentioned odd sector this was studied
systematically in \cite{Kampf:2011ty}.
Generally, the base or Lagrangian of the lowest lying resonances for $VVP$
gives complicated result with many parameters. Using the operator-product
expansion they are reduced just to two parameters. This verifies the so-called
LMD+P ansatz \cite{Moussallam:1994xp}.
For the on-shell pion we have only one parameter left and this can be set
using transition form factor. Another way how to set this parameter is to use
the information on $\rho\to\pi\gamma$ decay.  However, the experimental error of this value is
still big. Nevertheless one can use it as a consistency check
and it seems in good agreement \cite{Kampf:2011ty}.
The second parameter, which is connected with the off-shell pion can be obtained
from information on $\pi(1300)\to\rho \gamma$ and $\pi(1300)\to\gamma \gamma$.
However, here the experimental situation is even worse. Fortunately, there is at
least one experimental information from Belle: limit on $\pi(1300)\to\gamma
\gamma$ \cite{Abe:2006by}.

To summarize using the phenomenological information (from the space-like
region) we may set all relevant parameters within resonance chiral theory and
make some non-trivial predictions.


\subsection{Time-like region}
\noindent
In the time-like region the transition formfactor ${\cal
F}_{\pi^0\gamma\gamma}$ is mainly connected with the Dalitz decay $\pi^0\to
e^+e^-\gamma$ (see recent \cite{Kampf:2005tz} and references therein). For the
needs of the low-energy region it is convenient to study only a slope parameter
$a_\pi$:
\begin{equation}
 {\cal F}_{\pi^0\gamma\gamma^*}(q^2) = {\cal F}_{\pi^0\gamma\gamma^*}(0) \Bigl(
1+ a_\pi \frac{q^2}{m_{\pi^0}^2} + \ldots\Bigr)\,.
\end{equation}
Having the experimental data one should extract first the QED corrections:
\begin{equation}
\frac{\mathrm{d} \Gamma ^{exp}}{\mathrm{d}x}\,-\,
\delta_{QED}(x)\,\frac{\mathrm{d} \Gamma ^{LO}}{\mathrm{d}x}
\,=\,
\frac{\mathrm{d} \Gamma ^{LO}}{\mathrm{d}x}\,
[1 + 2x\, a_\pi]\,.
\end{equation}
These QED corrections are well understood \cite{Kampf:2005tz} and they include
now also one-photon irreducible contributions. Its value
\begin{equation}
 \delta a_\pi\Bigl|_\text{1$\gamma$\it IR} \doteq  0.005
\end{equation}
should be subtracted from the two relevant experiments \cite{slopy}. The
central value would shift to the excellent agreement with the theoretical
prediction
\begin{equation}
a_\pi^\text{theo} = 0.029 \pm 0.005\,.
\end{equation}
However, one should note that huge experimental errors (more than 100\%) make
the comparison meaningless.

\section{Conclusions}
\noindent
We have studied some basic properties of the pion formfactors at low energy.
We have first briefly discussed pion decay constant $F_\pi$ and set possible
inconsistency in this value obtained using $\pi_{\ell2}$ and $\pi^0\to
\gamma\gamma$. On the next object, electromagnetic formfactor of charged pion
we have demonstrated use of the so-called leading logarithms in studying the
convergence. Short overviews on scalar formfactor and radiative pion decay were
also given. Last but not least, $\pi-\gamma-\gamma$ transition formfactor was
discussed both for time-like and space-like region.

\section*{Acknowledgements}
\noindent
The work is supported by projects MSM0021620859 of Ministry of
Education of the Czech Republic.

\end{document}